# A Huffman based short message service compression technique using adjacent distance array

## Pranta Sarker*

Department of Computer Science and Engineering,
North East University Bangladesh,
Telihaor, Sheikhghat, Sylhet-3100, Bangladesh
Email: psarker@neub.edu.bd
*Corresponding author

## Mir Lutfur Rahman

Department of Computer Science,
University of Hertfordshire,
Hertfordshire, England, UK
Email: mirlutfur.rahman@gmail.com

**Abstract:** The short message service (SMS) is a wireless medium of transmission that allows you to send brief text messages. Cell phone devices have an uttermost SMS capacity of 1,120 bits in the traditional system. Moreover, the conventional SMS employs seven bits for each character, allowing the highest 160 characters for an SMS text message to be transmitted. This research demonstrated that an SMS message could contain more than 200 characters by representing around five bits each, introducing a data structure, namely, adjacent distance array (ADA) using the Huffman principle. Allowing the concept of lossless data compression technique, the proposed method of the research generates character's codeword utilising the standard Huffman. However, the ADA encodes the message by putting the ASCII value distances of all characters, and decoding performs by avoiding the whole Huffman tree traverse, which is the pivotal contribution of the research to develop an effective SMS compression technique for personal digital assistants (PDAs). The encoding and decoding processes have been discussed and contrasted with the conventional SMS text message system, where our proposed ADA technique performs outstandingly better from every aspect discovered after evaluating all outcomes.

**Biographical notes:** Pranta Sarker completed his Bachelor's in Computer Science and Engineering from North East University Bangladesh. He joined as a faculty member at the same university after graduation. He has been teaching for some great years as his passion. Currently, he is pursuing a Master's degree at Shahjalal University of Science and Technology, a preeminent computer science institution in Bangladesh. His research usually focuses on Huffman coding, data compression, Blockchain, and security.

Mir Lutfur Rahman also received his Bachelor's in Computer Science and Engineering from North East University Bangladesh. Then, he joined as a faculty member at the same university. Besides, his Master's degree is currently ongoing in Advanced Computer Science at the University of Hertfordshire. He has a good experience in teaching, and his research interest covers data compression, Huffman coding, computer vision, and machine learning.

# 1   Introduction

The Huffman principle serves as an essential strategy in the lossless data compression area by devising a minimal redundant coding scheme that allows each input symbol to have a variable code length tied to the symbol's occurrence (Huffman, 1952). This apposite algorithm, such as Huffman or arithmetic coding, generates a patchy distribution count, mainly when the characters nearby context is considered (Hossain et al., 2014). The dictionary-based compression methods are Ziv-Lempel (Fenwick, 1996), substituting the first occurrence of a string for the appearance of another exact string (Lelewer and Hirschberg, 1987). Today, Huffman coding approaches perform smooth data compression in data mining (Oswald et al., 2015) and wireless sensor networks (Săcăleanu et al., 2011; Renugadevi and Darisini, 2013) area. Recently, the short message service (SMS) texting and internet-formed short text messages (such as Messenger, WhatsApp, and Twitter) have sparked a surge in demand for short text message compression. In addition, mobile communication technology is such a convenient method that SMS, a text messaging technology, allows people to send and read messages with the absolute timing via mobile phones or personal digital assistants (PDAs) and personal computers in the form of instant messaging (IM).

However, now the question is, why do we compress short messages? Well, it is a fact that the compression of short text messages could save a significant bandwidth for a network if we have a large amount of SMS, Twitter, and IM traffic. Also, it could be possible to get an excellent outcome if we apply a standard compression technique for the large volume of data generated by buffering numerous messages during the real communication time, such as chat or IM (Kalajdzic et al., 2015). On the other hand, the one-time-password, a mobile phone-based two-factor authentication service over SMS, is getting more attention (Hallsteinsen et al., 2007; Eldefrawy et al., 2011). Besides, the maximum of 160 characters raises some limitations in the current system of SMS according to the standard GSM[1] 7-bit coding system, such as:

1   If we want to send a message with a combination of 165 characters, we have to require two SMS (160 + 5 characters) to send to the receiver, which is not a cost-effective way.

2   Most users strive to alter their message with subtly more than 160 characters into one SMS by getting back, deleting the 'unimportant' phases, or even changing the words as abbreviations as much as possible.

3    As a result, the unregulated use of abbreviations to save money by packing in single SMS resulted in many creepy messages that the receiver could only recognise if synchronised with the sender beforehand.

Likewise, it is worth noting that we have an inadequate supply of research about short text message compression and thus, considering the overall demanding circumstances, this research presents the *adjacent distance array* (ADA), a data structure formed on an array utilising the Huffman codes to compress SMS messages efficiently.

## 2    Related works

The Huffman data compression approach is utilised in several pieces of research. To illustrate, Musa et al. (2010) modified a compression mechanism to achieve high compression using the LZ-78 method for English text files while each character has appointed a weighted fixed code length resulting as an input in a binary file. In terms of text transformation, Senthil and Robert (2014) present a dictionary-based lossless reversible technique, namely, reinforced intelligent dictionary-based encoding that pre-processes the text and transforms it into an intermediate format, performs compression with more acceptable efficiency. In recent years, the block truncation coding, the least mean square error-based method, has been used as a lossy image compression mechanism (Chandravadhana and Nithiyanandam, 2017), which has lower computational complexity. To develop local language-based compression applications, (Gheraibia et al., 2018) discuss the effectiveness and compression rate based on entropy for a regional language known as the Cirebon language script employing Huffman Ternary code. Regarding the short text message compression, a Huffman table is constructed by Affandi et al. (2011) using the conventional Huffman tree to determine the codeword of SMS characters where it performs 28.73% in average compression ratio. (Mohd and Wong, 2012) presented an algorithm known as SMS zipper programmed with Java 2nd Micro Edition (J2ME) to develop mobile application software effective on Nokia S40 mobile series. For the Android system, Kumbhar and Krishnan (2012) adjusted arithmetic coding was applied to optimise most of the characters in the SMS body. An intelligent dictionary-based encoding utilises mobile phones' storage space in the present time with the best possible level of compression, space, time, overhead, and communication costs (Bhanarkar and Jha, 2012). *b64pack*, a novel method used to compress SMS according to Kalajdzic et al. (2015) to optimise with two such phases that outperform compress, gzip, and bzip2 in terms of the compression speed and ratio. The elliptic curve cryptographic encryption technique was utilised with compression in Patil et al. (2014) to increase the security of SMS transmission. Kirti et al. (2014) propose a user-friendly, cost-effective, and time-saving Huffman-based SMS compression technique consisting of character count and corresponding Huffman prefix code for each SMS character. A dynamic bit reduction algorithm implemented by Kaur and Garg (2015) is a key to finding the distinct characters to create a binary tree where the text compression and decompression happened using the modified Huffman method.

Although the ADA technique is implemented to achieve a significant encoding-decoding time and an excellent compression ratio for the English and transliterated Bengali language texts (Rahman et al., 2020; Sarker and Rahman, 2022), respectively; however, supported by the Huffman principle, this research introduces the method for encode-decode messages to remedy all of the issues associated with the SMS above. We have acquired a significant compression performance and bit reduction for each character. In addition, the proposed method can also send more than approximately 200 characters compared to the conventional GSM 7-bit encoding system, which utilises 160 characters by 7 bits each in PDU format according to the highest standard capacity is 1,120 bits for SMS.

## 3    Background

### 3.1   The SMS

Utilising several communication components such as mobile, web, or phone, the SMS follows a bi-directional communication by employing a standard short message peer-to-peer protocol that allows you to send text messages over a wireless network (Mahmoud et al., 2010; Kumbhar and Krishnan, 2012). An SMS centre (SMSC) allows a message to be stored first and then forwarded in place of sending SMS directly that keeps SMS until the target device is ready to get the message in the mobile telephone network. However, an SMS carries meta-data such as senders' information (name of the service centre and sender), protocol details (protocol identifier and data coding scheme), and timestamp. In the recipient's cell phone, the architecture for an SMS message's packets is defined as follows (Lo et al., 2008):

- *Header* means the type of message consisting of instruction sets such as air interface, SMSC, phone, and SIM card for the components of the SMS service network.

- *User data* indicates the body of the central message that the SMS sender wants to deliver, also known as payload.

GSM 7-bit, 8-bit, and 16-bit (UTF-16) data alphabet are the current three predominant encoding schemes for SMS (Kumbhar and Krishnan, 2012). According to the protocol data unit (PDU) mode, the standard GSM 7-bit alphabet data encoding scheme uses the Latin alphabet in most SMS messages (Croft and Olivier, 2005). PDU exchanges the information in several layers in the open system interconnection reference model and sends encoded messages over the GSM network (Bilgic and Sarikaya, 1992). Besides, 16-bit messages utilise 70 characters for non-Latin alphabets such as Chinese and Arabic in length (Le Bodic, 2005). However, this article pursues a practical approach to compressing the SMS body using the method of ADA based on the Huffman principle. Meanwhile, Table 1 presents the text limit for each corresponding SMS coding scheme as follows:

**Table 1**     Characters limit for each encoding system

| Encoding scheme | Maximum characters for each message segment | Text length |
|---|---|---|
| Traditional GSM 7-bit | 160 | 140 bytes or 1120 bits |
| 8-bit data | 140 | |
| 16-bit data alphabet | 70 | |

### 3.2 Formation of the SMS message

An SMS message might be up to 160 characters long (with every 7-bits of character), as endorsed by the European Telecommunications Standards Institute (Le Bodic, 2005). SMS text and SMS PDU are two modes stated in the SMS specification where the text mode is not accessible on all phones. However, any encoding can be used when PDU mode is employed. So instead of transmitting 7-bit American standard code for information interchange (ASCII) data, SMS PDU mode coded the information into a compressed form and transferred the octets represented in Table 2 as a fundamental structure (Ortiz and Prieto, 2004).

**Table 2** The elemental structure of the PDU packet

| Name | SCA | PDUT | MR | DA | PID |
|---|---|---|---|---|---|
| Defines | Service center | Protocol data unit | Message reference | Destination address | Protocol identifier |
| Name | DCS | VP | | UDL | UD-PDU |
| Defines | Destination address | Validity period | | User data length | User data PDU |

### 3.3 The concept of Huffman coding

Huffman's theory is more sophisticatedly considered an asset in the lossless data compression mechanism. Forming by the occurrence of characters in a text, it erects a binary tree. Figure 1 illustrates a Huffman tree for a frequency table in Table 3. The entire process follows a minimum-redundancy coding technique that defines a character getting the longest codeword (bit sequences) for the minimum occurrence and the lowest codeword for the maximum appearance (Huffman, 1952). That variable-length codeword represents each character with a unique prefix code, also known as prefix-free codes (Mohd and Wong, 2012; Patil et al., 2014) to eliminate ambiguity while decoding the codewords. Nevertheless, the construction system of binary Huffman is consecutive and organised, albeit the tree has a balancing issue (Rajput, 2017; Habib et al., 2018). Each node preserves the relative occurrence probability in the subtree underneath the node of the characters. In addition, the leaf nodes indicate original characters encoded with the binary bits zero and one assigned to the edges.

**Figure 1** Construction of regular Huffman tree (see online version for colours)

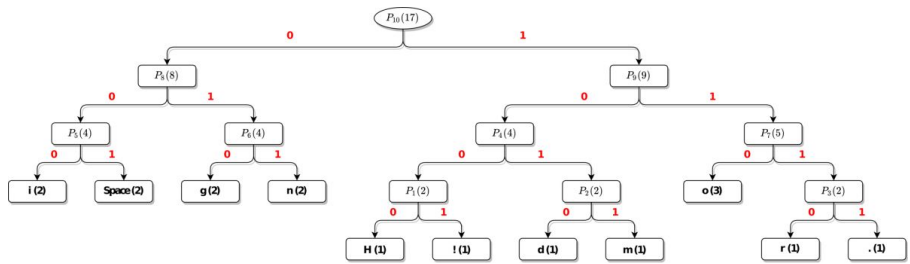

**Table 3**     Frequency table generated by binary Huffman coding

| Character | Frequency | Codeword |
|-----------|-----------|----------|
| o | 3 | 110 |
| g | 2 | 010 |
| n | 2 | 011 |
| i | 2 | 000 |
| <space> | 2 | 001 |
| H | 1 | 1000 |
| ! | 1 | 1001 |
| d | 1 | 1010 |
| m | 1 | 1011 |
| r | 1 | 1110 |
| . | 1 | 1111 |

## 4   Architecture

### 4.1   Codeword generation for each character

We construct a frequency table to generate a codeword for each character compliant with the binary Huffman technique. The table lists every distinct character and its frequency for the entire SMS message, where each character has a unique variable-length bit string, $L$ in Table 3, for a meaningful SMS text message, $M$ = 'Hi! good morning.'

To construct the Huffman binary tree, we have considered a set of characters $C = \{c_0, c_1, \ldots, c_{n-1}\} = \{o, g, \ldots, .\}$ with frequencies $F = \{f_0, f_1, \ldots, f_{n-1}\} = \{3, 2, \ldots, 1\}$ for $f_0 \geq f_1 \geq \ldots \geq f_{n-1}$ where each character $c_i$ has a frequency $f_i$ and $n$ represents the number of characters. For each character $c_i$, $0 \leq i \leq n - 1$, traversing a path from root preserves the codeword $w_i$ for that particular character. The journey writes '0' and '1' to the left and right child in the tree, producing a set of distinct codewords $W = \{w_0, w_1, \ldots, w_{10}\} = \{110, 010, \ldots, 1111\}$ for Figure 1. The level of codeword could be determined according to the level of $c_i$ if the root level is zero, and the minimum expected weighted path length $\sum f_i l_i$ resolute the traversing time of the tree.

### 4.2   The proposed methodology

In the concept of *adjacent*, we are considering the immediate character of the SMS text message from left to right. For instance, if we have a text message 'good morning', the character 'g' has an adjacent character is 'o', and 'd' and so on. To describe the *Distance*, we could execute a subtract operation between two ASCII symbols, such as English letters 'a' and 'c' are derived as 97 and 99 in the ASCII character set (Gottfried and Chhabra, no date). Therefore, the distance between them would always be two, whatever we choose from 'a' to 'c' → (–2) or 'c' or 'a' → (2). On the other hand, the *Array*, a linear data structure, conserves the distance values of adjoining characters.

### 4.2.1 Compression using ADA

This research endorsed the array of adjacent characters, denoted by, *adjacent* that operates the encoding and decoding procedures. However, a threshold value, *T* would be utilised to identify the adjacent characters that assist in alleviating a prolonged code block generated from the encoded file, denoted by *encoded* for each adjacent. Moreover, for a specific symbol $S_i$, the adjacent symbols would be $S_{i+1} + S_{i+2} + \ldots + S_{i+m}$ if the (1) is shown, explaining an absolute ASCII value distance between two symbols, such as $S_i$ and $S_{i+1}$ whether it is less than or equivalent to the *T*.

$$T \geq \left| S_{i+1} - S_i \right| \tag{1}$$

The case distinctly allows for each $S_i$,

• The *encoded* to put the codeword generated from the Huffman principle.

• The *adjacent* to keep distances for all fulfilled characters.

Otherwise,

• $S_i$ would be the new symbol in *encoded*.

The overall process would be repeated until we reach the message's conclusion. Nevertheless, the *encoded* obtain a bit '0' as a 'separator' to the disjunction between two distinct symbols. Depending on the threshold value, *T*, the distances of adjacent symbols are delineated by a particular two distinct coding schemes in the *adjacent*.

• The first has 2-bit patterns, such as '$1\{0/1\}\cdots$' where,

    1  The first bit, '1' defines the commencing of the code, and

    2  The second bit recognises whether the distance value is positive or negative. In that case, we have considered '0' and '1' as positive and negative values, respectively.

• Rest is the binary code for the adjacent distance of symbols.

Meanwhile, we might calculate the threshold value in equation (2) since binary codes of adjacent distances for each symbol has a commensurate bit length of *T* that could be maximum by:

$$T = 2^x - 1 \tag{2}$$

where *x* presents the most significant numeral bits for threshold value, *T* to perform accurately for our technique. Therefore, the memory requirement for the *encoded* is:

$$\delta_1 = \sum_{i=1}^{N} \left( F_i - A_i \right) \times C_i \tag{3}$$

Equation (3) represents:

$F_i$   the number of occurrences or frequency of a character.

$A_i$   from the *encoded* into *adjacent* the abatement of occurrences.

$C_i$   Huffman generated codeword's bits.

*N*    Total symbols or characters in the *encoded*.

Furthermore, *adjacent* allocates the memory as:

$$\delta_2 = \sum_{i=1}^{M} A_i \times x \tag{4}$$

Equation (4) represents:

*x*    bit amount for the distances.

*M*    total distances for adjacent characters.

Hence, we need to represent the memory for an SMS message is computed by:

$$\delta = \delta_1 + \delta_2 + H_T + S_N \tag{5}$$

In equation (5), for the Huffman tree, $H_T$ represents its header and $S_N$ is the total separators for each distinct character in *adjacent*. Using the Huffman principle, our approach allows a dictionary to perform a compression process for the complete SMS message *M* in Figure 2. In this dictionary, the first line defines the total number of unique symbols, and the rest of the lines have the character, the length of a codeword, and the ultimate codewords.

**Figure 2**    The dictionary is created for *M* (see online version for colours)

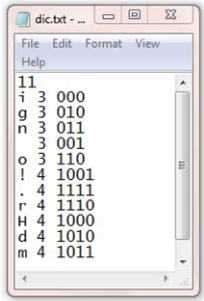

**Figure 3**    The encoded file (see online version for colours)

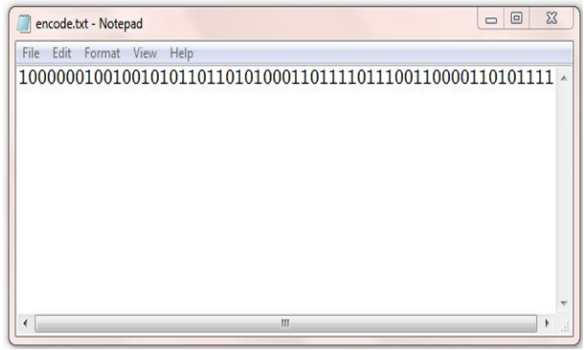

To generate the encoded file manifested in Figure 3, we have assumed a threshold value $T = 7$, which means the maximum number of bits for the $T$ have $x = 3$. Considering the $M$, we have found the *adjacent* in Figure 4 with a full of separator bit of zeros because that could not meet equation (1) for the first letter 'H' in the message. As a result, every character of the SMS message is encoded by their codeword in the *encoded* for Figure 3, such as 1,000 (first four bits) represents the character 'H', 000 (next three bits) represents 'i' on the entire file we have 1,001 for '!', 001 for <space>, 010 for 'g', 110 for 'o', 1010 for 'd', 1011 for 'm', 1110 for 'r', 011 for 'n', and 1111 for '.'.

**Figure 4**    The ADA file with the coding scheme (see online version for colours)

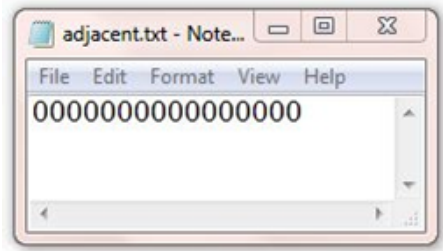

### 4.2.2 Decompression using ADA

The decompression process of our method allows a decoded file that decodes the initial character (such as $S_i$) from *encoded*. Then, computing the distances from the *adjacent* for that character, the consecutive symbols (such as $S_{i+1}$, $S_{i+2}$ + … +$S_{i+m}$) would be decoded. The operation persists since a bit separator '0' is introduced or the *adjacent* reaches its limit. The method neither allows traversing the whole Huffman code list nor the entire Huffman tree. Since we have found '0' as separator bits in the *adjacent* and codewords for all characters in the *encoded*; as a result, we could comfortably decode the SMS message in Figure 5 that has been taken into account above.

**Figure 5**    The decoded version of the SMS message (see online version for colours)

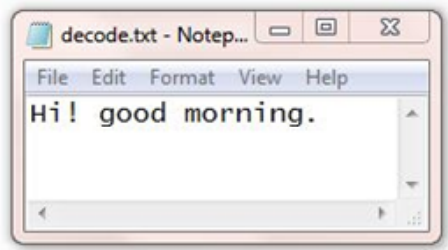

## 5    Implementation

### 5.1   *Analysis of encoding process*

Before beginning the process, we have an SMS message *M*, a specific *T* and the ultimate *L*. Meanwhile, for each symbol $S_i$ of *M*, the process would justify two different conditions from a particular symbol $S_i$.

1    The threshold value *T* should be greater or equal to the adjacent distances (such as $D_{i+1}$, $D_{i+2}$, …, $D_{i+m}$) for the adjacent symbols (such as $S_{i+1}$, $S_{i+2}$, …, $S_{i+m}$).

2    The Huffman generated codeword's bit length (such as $L_{i+1}$, $L_{i+2}$, …, $L_{i+m}$) should be greater or equal to the bit length of each adjacent symbol.

**Figure 6**    A working flow for the encoding process (see online version for colours)

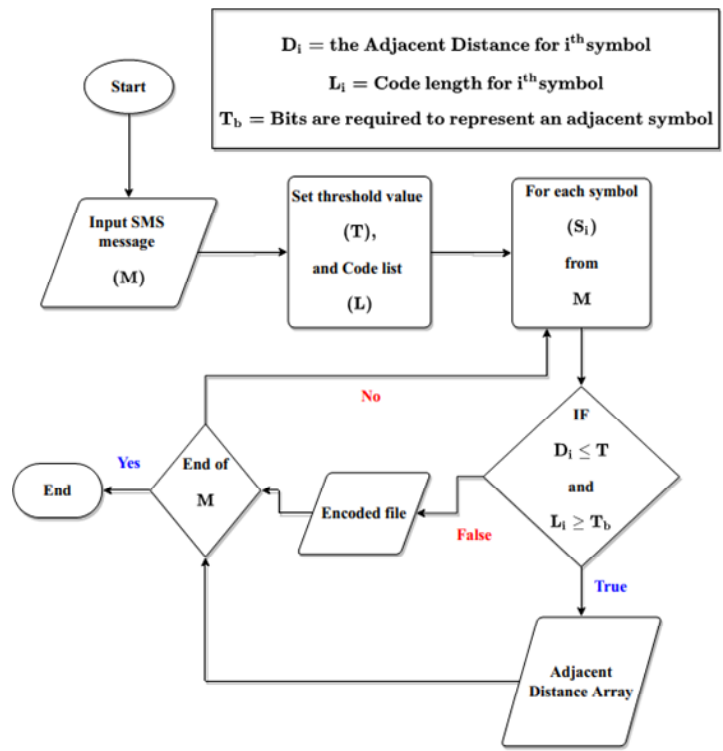

Where *m* = *l* – 1, and *l* is the length of *M*. The symbol $S_i$ would be saved into *encoded* as a new symbol for the mismatching of views (1 and 2) above; otherwise, the *adjacent* would keep the distances (such as $D_{i+1}$, $D_{i+2}$, …, $D_{i+m}$) and allow separator bit '0' after saving all distances from the *encoded*. The entire procedure is lengthy until to find end of *M*. Figure 6 provides the mechanism of the whole process.

Therefore, the process allows us to get two several files; moreover, the coded files such as:

- the encoded coded file is similar to Figure 3

- the adjacent coded file is similar to Figure 4.

## 5.2 *Analysis of decoding process*

The decoding process utilises the files *encoded*, *adjacent*, and the *L*. The process decodes the earliest symbol (such as $S_i$) from the file *encoded*. Subsequently, the rest of the adjoining symbols (such as $S_{i+1}$, $S_{i+2}$, …, $S_{i+m}$) could be decompressed by calculating the corresponding ASCII value distances considering the file *adjacent*. The process would work until:

- To get '0', a separator bit in the file, *adjacent*.

- To reach the final destination of the *adjacent*.

The analysed mechanism of the decoding process provides in Figure 7.

**Figure 7**  A working flow for the decoding process (see online version for colours)

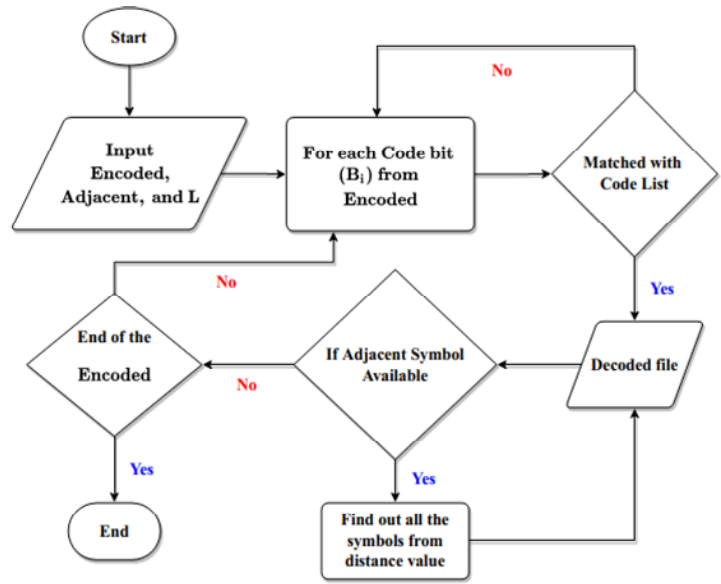

## 6   Results and discussion

The section assesses our proposed methodology's performance, contrasting the traditional GSM 7-bit alphabet encoding system for the SMS. Since we have a limitation of standard corpus upon SMS messages, we have esteemed a renowned SMS text message corpus based on arbitrary topics of public conversations in English and Mandarin Chinese in English (Chen and Kan, 2012). To evaluate the desired outcome of our method, we have classified the sample messages into five test cases among 71,000 or more messages. As an SMS text message belongs to a short message sent from one to another, in that case, only for the assessment purpose, we have taken five consecutive sample messages from the corpus for each test case illustrated in Table 4, where five consecutive messages have each to determine the performance of our method. Note that there have two attributes for each message, such as *message-id*, denoted by *<message id = 's'>*, where *s* is the consecutive serial number for each message, and the ultimate text message, presented by *<text> M </text>*, where *M* represents the actual message of the SMS.

The experiment ran on the Microsoft 64-bit operating system (Windows 7) with hardware such as Primary Memory – 3.88 GiB, the Processor profile is Intel(R) Core (TM) i3-4160 CPU @ 3.60 GHz x 4, Intel(R) HD Graphics 4400 (memory 1696 MB) – Graphics. Moreover, the relevant compiled codecs were in the GCC version 8.1.0 system.

**Table 4**    The organisation of test cases for the SMS message

| Test cases | SMS text messages | Total length |
|---|---|---|
| T1 | <message id="10120"><text>Bugis oso near wat...</text> <message id="10121"><text>Go until jurong point, crazy.. Available only in bugis n great world la e buffet... Cine there got amore wat...</text> <message id="10122"><text>I dunno until when... Lets go learn pilates...</text> <message id="10123"><text>Den only weekdays got special price... Haiz... Cant eat liao... Cut nails oso muz wait until i finish drivin wat, lunch still muz eat wat...</text> <message id="10124"><text>Meet after lunch la...</text> | $21 + 111 + 46 + 140 + 22 = 340$ |
| T2 | <message id="10125"><text>m walking in citylink now u faster come down... Me very hungry...</text> <message id="10126"><text>5 nights...We nt staying at port step liao...Too ex</text> <message id="10127"><text>Hey pple...D700 or D900 for 5 nights...Excellent location wif breakfast hamper!!!</text> <message id="10128"><text>Yun ah.the ubi one say if u wan call by tomorrow.call 67441233 look for irene.ere only got bus8,22,65,61,66,382. Ubi cres,ubi tech park.6ph for 1st 5wkg days.en</text> <message id="10129"><text>Hey tmr maybe can meet you at yck</text> | $65 + 51 + 81 + 160 + 33 = 390$ |
| T3 | <message id="10130"><text>Oh...i asked for fun. Haha...take care. u</text> <message id="10131"><text>We are supposed to meet to discuss abt our trip... Thought xuhui told you? In the afternoon. Thought we can go for lesson after that</text> <message id="10132"><text>t finish my film yet...</text> <message id="10133"><text>m having dinner with my cousin...</text> <message id="10134"><text>Oh... Kay... On sat right?</text> | $41 + 132 + 23 + 33 + 26 = 255$ |

**Table 4**    The organisation of test cases for the SMS message (continued)

| Test cases | SMS text messages | Total length |
|---|---|---|
| T4 | <message id="10135"><text>I need... Coz i never go before</text> <message id="10136"><text>s a basic yoga course... at bugis... We can go for that... Pilates intro next sat.... Tell me what time you r free</text> <message id="10137"><text>I am going to sao mu today. Will be done only at 12</text <message id="10138"><text>Hey gals...U all wanna meet 4 dinner at nite?</text> <message id="10139"><text>Jos ask if u wana meet up?</text> | 31 + 114 + 51 + 45 + 26 = 267 |
| T5 | <message id="10140"><text>Haiyoh... Maybe your hamster was jealous of million</text> <message id="10141"><text>is your hamster dead? Hey so tmr i meet you at 1pm orchard mrt?</text> <message id="10142"><text>ve booked the pilates and yoga lesson already... Haha</text> <message id="10143"><text>Yup... I havent been there before... You want to go for the yoga? I can call up to book</text> <message id="10144"><text>K... Must book a not huh? so going for yoga basic on sunday?</text> | 51 + 63 + 53 + 87 + 60 = 314 |

### 6.1 Compression result of our proposed technique

The experimental result of our method is shown in Table 5 for each test case considering two threshold values $T = 7$ and $T = 15$. To evaluate the compression performance, it would cover the consolidated result of bits for two coded files as we described earlier in Section 5.1, such as:

- adjacent.txt – for the *adjacent*
- encode.txt – for the *encoded*.

However, the resulting analogous data for the traditional GSM 7-bit encoding system explicitly indicates an inferior compression performance to our proposed approach. Moreover, the ADA technique representing each character by around 5-bits is significantly improved than the conventional GSM system.

### 6.2 Maximum number of characters of our method

Characters or symbols are essential to writing an SMS to send the message to the receiver, and the meaning of the SMS should be legible so that the receiver can easily recognise it. The conventional 7-bit GSM alphabet SMS encoding system allows separate messages for more than its limit of 160 characters (Husodo and Munir, 2011; Moses and Deisy, 2014). Nevertheless, the proposed approach affords more than 200 characters and achieves a notable character enhancement for threshold value $T = 7$, and $T = 15$ displayed in Table 6.

**Table 5**    The comparison of compression performance

| Test cases | The total length for each test case ($L_{x_n}$) | Compliant with the proposed method for | Aggregated number of bits for proposed method (encode.txt + adjacent.txt) ($TB_{x_n}$) | Required bits to represent a character for the proposed method ($\frac{TB_{x_n}}{L_{x_n}}$) | Aggregated number of bits for the traditional GSM SMS system ($TB_n$) | Required bits to represent a character for the conventional GSM SMS system ($\frac{TB_n}{L_{x_n}}$) |
|---|---|---|---|---|---|---|
| T1, n = 1 | 340 | $X = T = 7$ | $1344 + 431 = 1775$ | 5.22 | 2380 | 7.00 |
|  |  | $X = T = 15$ | $1308 + 474 = 1782$ | 5.24 |  |  |
| T2, n + 2 | 390 | $X = T = 7$ | $1463 + 657 = 2120$ | 5.43 | 2730 |  |
|  |  | $X = T = 15$ | $1448 + 714 = 2162$ | 5.54 |  |  |
| T3, n = 3 | 255 | $X = T = 7$ | $951 + 370 = 1321$ | 5.18 | 1785 |  |
|  |  | $X = T = 15$ | $997 + 344 = 1341$ | 5.25 |  |  |
| T4, n = 4 | 267 | $X = T = 7$ | $1001 + 378 = 1379$ | 5.16 | 1869 |  |
|  |  | $X = T = 15$ | $1050 + 341 = 1391$ | 5.20 |  |  |
| T5, n = 5 | 314 | $X = T = 7$ | $1110 + 513 = 1623$ | 5.16 | 2198 |  |
|  |  | $X = T = 15$ | $1227 + 433 = 1660$ | 5.28 |  |  |

**Table 6**     Highest adjustment and enhancement of characters

| Test cases | $L_{T_n}$ | Compliant with the proposed method for | $TB_{X_n}$ | Maximum characters adjustment $Mxchar_{X_n} = \left(1120 \div \dfrac{TB_{X_n}}{L_{T_n}}\right)$ | Enhancement of characters (in %) $\left(\dfrac{Mxchar_{X_n}-160}{160}\right)\times100\%$ |
|---|---|---|---|---|---|
| T1, | 340 | $X = T = 7$ | 1775 | 214.54 | 34.09 |
| $n = 1$ | | $X = T = 15$ | 1782 | 213.69 | 33.56 |
| T2, | 390 | $X = T = 7$ | 2120 | 206.04 | 28.78 |
| $n = 2$ | | $X = T = 15$ | 2162 | 202.04 | 26.28 |
| T3, | 255 | $X = T = 7$ | 1321 | 216.20 | 35.13 |
| $n = 3$ | | $X = T = 15$ | 1341 | 212.98 | 33.11 |
| T4, | 267 | $X = T = 7$ | 1379 | 216.85 | 35.53 |
| $n = 4$ | | $X = T = 15$ | 1391 | 214.98 | 34.36 |
| T5, | 314 | $X = T = 7$ | 1623 | 216.69 | 35.43 |
| $n = 5$ | | $X = T = 15$ | 1660 | 211.86 | 32.41 |

Albeit the impaction is nominal after estimating all analogous data, there has been an observation regarding the performance that we could get an effective and optimum performance for the threshold value $T = 7$ rather than $T = 15$ in terms of such parameters as utilising bits, adjusting the maximum, and enhancement of characters for each test cases. Meanwhile, the overall performance is juxtaposed with the conventional GSM 7-bit SMS encoding process summarised in a graph in Figure 8 and Figure 9 for utilising total bits and maximum characters, reflecting an explicit and outstanding refinement of the proposed methodology.

**Figure 8**     Refinement of using bits for the proposed method compared to the traditional GSM 7-bit (see online version for colours)

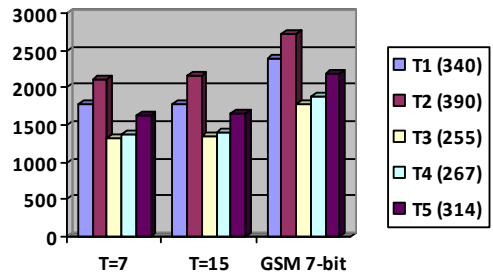

The experimental study perceives a better encoding performance while the threshold value finds minimum as possible, indicating the fewer adjacent characters. On the other hand, we could get many adjacent symbols in the *adjacent* for maximising the threshold value, which grows the size of the *encoded*. As a result, the threshold value is an attribute that works as a catalyst to tune the compression performance of the SMS message. There is some concern about the rising threshold values; however, we aim to minimise the threshold value to get the optimum outcome. Therefore, our research based on the ADA methodology has tremendous potential to replace the traditional GSM 7-bit SMS system in a broad sense. Furthermore, it could be a footprint by contributing knowledge to the SMS text compression study.

**Figure 9**    Refinement of adjusting the maximum characters for the proposed method contrasted to the traditional GSM 7-bit (see online version for colours)

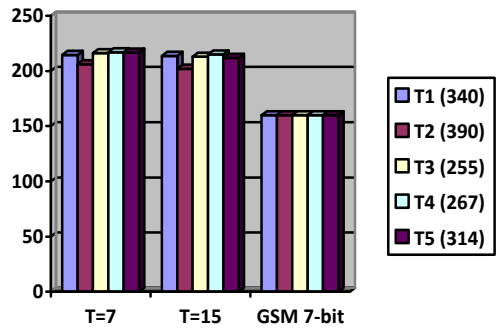

## 7    Conclusions and future work

The study presents an array data structure, which keeps the distances of adjacent symbols, known as ADA, utilising the Huffman codes to perform compression upon SMS text messages. As a part of the lossless data compression mechanism, our research uses the binary Huffman concept to produce the character's codeword. After that, the threshold parameter and ADA manage the whole methodology, such as the process does not need to traverse any tree during the decoding. The encoding process of the study produces two coded files by tuning the threshold values. We have gained an outstanding experimental study upon some test cases of famous corpus for assessment parameters such as total using bits, bits per character, maximum adjustment of characters, and the enhancement of characters (in %). Our proposed ADA method outperforms the 7-bit GSM default alphabet encoding system from every assessment aspect. The figure is about fewer characters for each test case, around 5-bits per character, and almost 33.79% average enhancement of more than 200 characters to construct a message for $T = 7$, whereas the existing GSM 7-bit SMS limits only 160 characters holding 7-bit each for a message. Although we expect better performance for the proposed technique by updating the experimental environment; however, the major issue of the higher threshold value and the growing adjacent characters could be invested in future research.

## Notes

1    Global system for mobile communications (GSM) is a benchmark for cellular networks.